\documentclass[twocolumn,showpacs,prl,preprintnumbers,superscriptaddress]{revtex4-1}

\usepackage{color, amsmath, amssymb, graphicx, float, dcolumn, bm}

\usepackage{revsymb,natbib,placeins,setspace}

\usepackage[dvipsnames]{xcolor}

\usepackage{epstopdf}

\usepackage{etoolbox}

\begin{document}

\title{Global network control from local information}

\author{Aleksandar Haber} 
\affiliation{Department of Physics and Astronomy, Northwestern University, Evanston, IL 60208, USA} 

\author{Ferenc Molnar} 
\affiliation{Department of Physics and Astronomy, Northwestern University, Evanston, IL 60208, USA}

\author{Adilson E. Motter}
\affiliation{Department of Physics and Astronomy, Northwestern University, Evanston, IL 60208, USA}
\affiliation{ Center for Network Dynamics, Northwestern University, Evanston, IL 60208, USA}
\affiliation{Northwestern Institute on Complex Systems, Northwestern University, Evanston, IL 60208, USA}
\affiliation{Department of Engineering Sciences and Applied Mathematics, Northwestern University, Evanston, IL 60208, USA}

\begin{abstract}
In the classical control of network systems, the control actions on a node are determined as a function of the states of all nodes in the network. Motivated by applications where the global state cannot be reconstructed in real time due to limitations in the collection, communication, and processing of data, here we introduce a control approach in which the control actions can be computed as a function of the states of the nodes within a limited {\it state information neighborhood}. The trade-off between the control performance and the size of this neighborhood is primarily determined by the condition number of the controllability Gramian. Our theoretical results are supported by simulations on regular and random networks and are further illustrated by an application to the control of power-grid synchronization. We demonstrate that for well-conditioned Gramians, there is no significant loss of control performance as the size of the state information neighborhood is reduced, allowing efficient control of large networks using only local information.  
\end{abstract}

\maketitle  

\noindent
{\bf The study of dynamical processes on networks has been furthered by the prospect that many such processes can be controlled. An outstanding challenge is that large networks of current interest have high-dimensional state spaces and traditional control approaches are not scalable to high dimensions. A key factor limiting scalability is the information needed to compute the control actions at each controlled node of the network. In this study, we introduce a non-iterative, closed-form solution for network control that only requires local information from a select set of nodes. The approach is based on the notion of {\it state information neighborhood} and thus constitutes an open-loop counterpart of a local approach recently introduced for feedback control \cite{duan2022}.  A notable attribute of the approach is its ability to systematically and accurately determine both the number and the identity of the nodes in the state information neighborhoods of all nodes. The results show that large networks can be controlled using state information from small neighborhoods.}

\section{Introduction}

The problem of controlling network dynamics is important in many applications, 
such as  in mitigating cascading failures in power systems~\cite{anderson2008power}, 
identifying drug targets in genetic networks~\cite{zanudoPLOS2015,wellsPRX2015},  
designing ecosystem management approaches~\cite{lessard2005should,sagar2011},
and coordinating the distributed control of robotic networks~\cite{bullo2009distributed}, 
and has received significant attention in the recent physics literature~\cite{scholl,reviewLiu,reviewMotter}. 
This attention stems from the systems-level nature of the control principles and the general notion that these principles can provide answers to fundamental questions on complex systems that cannot be addressed by traditional research. From a theoretical standpoint, the problem is to find a control input that will drive the network from a given initial state to a desired state. In the synchronization of power-grid networks~\cite{rohdenPRL2012,motter2013spontaneous,bulloPNAS2013}, for example, the initial state may correspond to a post-perturbation scenario in which the power generators would eventually desynchronize, thereby compromising their operation, while the desired state corresponds to a stable synchronous state.

In recent approaches for controlling networks, the control inputs are often expressed in terms of the inverse of the
controllability Gramian  
(e.g.,~\cite{gao2014target,motter2013,bulloIEEE2014,Gang2016}).
However, even when the dynamics are provably controllable, this formulation can
fail numerically in large networks because the inversion of the Gramian can be
computationally infeasible~\cite{motter2013,haber2016}. Moreover,  such 
approaches seek to determine the control inputs for a node as a function of the initial and desired states of all nodes in the network. In practice, a controlled node often can only assess the state information of nodes that are within a certain neighborhood in the network. 
Even when state information is not limited, computing the control actions as functions of the whole network state can be computationally prohibitive.
These difficulties motivate us to consider the following questions:  To what extent can the control actions for a node be determined on the basis of the state of its local neighborhood?
How does the control performance depend on the 
size of this neighborhood, the controllability Gramian, and the properties of the network?  

In this article, we show that in networks whose first few powers of the adjacency matrices are 
sparse---which include many regular, random, and real networks---the control actions on a node can be determined by only using the initial and desired state information  
in a 
neighborhood of the node. These neighborhoods will be referred to as 
 \textit{state information neighborhoods} (SINs).
Our approach does not require explicit inversion of the controllability Gramian and is thus applicable even when the Gramian is ill-conditioned \cite{motter2013}. 
We show, however, that for fixed SINs
the control performance improves as the condition number of the Gramian decreases.
Moreover, for 
well-conditioned Gramians, 
we can choose very small SINs
without incurring a significant loss of control performance. 
Our approach yields a closed-form solution that does not rely on iterative procedures that might suffer from convergence and implementation issues; unlike other closed-form decentralized 
and distributed methods~\cite{Siljak12,lin13},  our control approach
scales well (in fact {\it linearly}) with network size.

An important aspect of our approach is that it only requires communication {\it within} the local SINs. 
The approach leverages locality inherited from the sparsity and/or off-diagonal decay of the relevant matrices in a broad class of network systems, which is reflected in the solutions of various equations pertinent to control, including the Riccati and Lyapunov equations \cite{bamieh2002,motee2008,curtain2011}.
Significant optimization and state estimation approaches that benefit from sparsity have been previously proposed, including approaches for nonlinear constrained optimization \cite{hespanha2022}, distributed implementation of Kalman filters \cite{khan2008}, and state estimation problems in distributed settings \cite{pasqualetti2012}. These approaches are iterative and generally involve network communication or consensus. We reiterate that, in contrast, our approach is non-iterative (with calculations implemented in a single step) and avoids large-scale communication,  which favors scalability. The approach is thus an open-loop counterpart of a local approach recently introduced in Ref.~\cite{duan2022} for feedback control.

\section{Localized Control Approach}

We consider networks consisting of $N$ nodes  described by the state equation
$\mathbf{x}(k+1)=A\mathbf{x}(k)+B\mathbf{u}(k)$, 
where
the dynamics of the nodes is  $n$ dimensional,
$\mathbf{x}(k)=\text{col}\left(\mathbf{x}_{1}(k),\mathbf{x}_{2}(k),\ldots,\mathbf{x}_{N}(k)\right)\in\mathbb{R}^{Nn}$ is the network state at the discrete time $k$,  
$\mathbf{u}(k)=\text{col}\left(\mathbf{u}_{1}(k),\mathbf{u}_{2}(k),\ldots,\mathbf{u}_{N}(k)\right)\in\mathbb{R}^{Nm}$ is the control input,  $A\in\mathbb{R}^{Nn\times Nn}$ is the coupling matrix~\cite{netgen}, 
and $B \in \mathbb{R}^{Nn \times Nm}$ is a block diagonal input matrix formed by $n\times m$ blocks.
The vectors $\mathbf{x}_{i}(k)\in\mathbb{R}^{n}$ and $\mathbf{u}_{i}(k)\in\mathbb{R}^{m}$ are, respectively, the state and control input of
node $i$. If certain nodes or local state variables are not controlled, then the corresponding entries of $B$ are zero. Starting with $k=0$ and  propagating the dynamics
for $f$ time steps, we obtain
\begin{align}
\mathbf{x}(f)=\mathcal{A}\mathbf{x}(0)+K_{f}\underline{\mathbf{u}}^{f},
\label{leastSquaresFormulation}
\end{align}
where  $\mathcal{A}=A^{f}$, $K_{f}=\begin{bmatrix}A^{f-1}B&A^{f-2}B&\ldots&AB&B\end{bmatrix}\in\mathbb{R}^{Nn\times Nfm}$ is the \textit{$f$-step controllability matrix},
and $\underline{\mathbf{u}}^{f}=\text{col}\left(\mathbf{u}(0),\mathbf{u}(1),\ldots,\mathbf{u}(f-1)\right)\in\mathbb{R}^{Nfm}$ is the \textit{input sequence vector}.
Equation~\eqref{leastSquaresFormulation} has an elegant graph interpretation as it shows that the state $\mathbf{x}_{i}$ of node $i$ at time $f$ is influenced by the states at time $0$ of all nodes that are within distance $f$  from node $i$ in the network---further details are presented in Supplementary Material, Sec.~S2. In the simulations below, unless otherwise noted, 
we use $m=n=3$.

In our formulation, the control objective 
is to drive the network from the initial state $\mathbf{x}(0)$ to a desired state $\mathbf{x}^d$ in $f$ time steps by applying the input sequence  $\underline{\mathbf{u}}^{f}$. The control problem
is thus to find $\underline{\mathbf{u}}^{f}$ by solving Eq.~\eqref{leastSquaresFormulation}, under the condition that $\mathbf{x}(f)=\mathbf{x}^{d}$. 
In the following, we assume that $A^{l}$ are sparse for all $l=1, \dots , 2f-1$, which 
ensures that the matrices $\mathcal{A}$ and $K_{f}$ are sparse.
To guarantee the existence of the exact solution, we would also
need to assume that $f\ge\nu$, where  $\nu$  is the smallest integer such that $K_{\nu}$ is full 
rank~\cite{luenberger1971} (we anticipate that this condition can be relaxed within our approach below). 
To proceed, we permute the entries of vector $\underline{\mathbf{u}}^{f}$~\cite{haber2014}. First, for every node $i$, we define its input sequence vector $\overline{\mathbf{u}}_{i}^{f}=\text{col}\left(\mathbf{u}_{i}(0),\mathbf{u}_{i}(1), \ldots,\mathbf{u}_{i}(f-1)\right)$. Then, we define the vector $\overline{\mathbf{u}}^{f}=\text{col}(\overline{\mathbf{u}}_{1}^{f},\overline{\mathbf{u}}_{2}^{f}, \ldots ,\overline{\mathbf{u}}_{N}^{f})$. The definition of $\overline{\mathbf{u}}^{f}$ induces a permutation matrix $P\in\mathbb{R}^{Nm\times Nm}$ such that $\overline{\mathbf{u}}^{f}=P\underline{\mathbf{u}}^{f}$. From $P^{T}P=I$ and Eq.~\eqref{leastSquaresFormulation}, we have
\begin{align}
 \mathbf{x}(f) =& \mathcal{A}\mathbf{x}(0)+\mathcal{K}_{f}\overline{\mathbf{u}}^{f} \label{leastSquaresFormulation2},   
\end{align}
where $\mathcal{K}_{f}=K_{f}P^{T}$. 
 The minimum-energy input sequence that solves Eq.~\eqref{leastSquaresFormulation2} is determined as the solution~of 
$ \min \left\|\overline{\mathbf{u}}^{f} \right\|_{2}^{2}$ subject to $\mathbf{x}^{d} =\mathcal{A}\mathbf{x}(0)+\mathcal{K}_{f}\overline{\mathbf{u}}^{f}$,
where $\left\|\cdot \right\|_{2}$ is
the $2-$norm~\cite{rugh1996linear}.  
This solution can be expressed as
\begin{align}
\overline{\mathbf{u}}^{f}=\mathcal{K}^{T}_{f}W^{-1}_{f}\left[\mathbf{x}^{d}  -\mathcal{A}\mathbf{x}(0) \right],  
\label{solutionMinNorm}
\end{align}
where $W_{f}=\mathcal{K}_{f}\mathcal{K}^{T}_{f}=\sum_{j=0}^{f-1}A^{j}BB^{T}(A^{T})^{j}\in \mathbb{R}^{Nn \times Nn}$ is the \textit{controllability Gramian};
 the condition $f\ge \nu$  guarantees the existence of $W^{-1}_{f}$. 
We note that, in addition to the minimal-energy formulation, the theory we present 
also applies to 
other optimal feedforward control formulations (Supplementary Material, Sec.~S8).

\begin{figure}[t!] 
\centering
\includegraphics[scale=1]{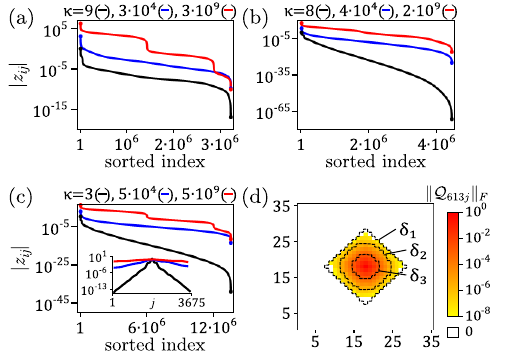}
\caption{Ordered magnitudes of the entries of $W_f^{-1}$
for (a) an  ERN ($N=598$), (b) an RGN ($N=702$), and  (c) a lattice  ($N=35^2$), for $B=B_1$ (black), $B_2$ (blue), and $B_3$ (red). Inset in (c): corresponding envelope of the off-diagonal decay for the middle row as a function of the column. (d) SIN and effective SINs for the matrix $\mathcal{Q}$ around a central node for the $B=B_{1}$ case in (c) and thresholds $\delta_{1}=0$, $\delta_{2}=10^{-6}$, and $\delta_{3}=10^{-3}$.   Other parameters: $f=5$ in (a-d) and $q=3$ in (d)~\cite{param}. Details on 
network generation, sparsity of $W_{f}$, localization of $W_{f}^{-1}$, and (effective) SINs are presented in Supplementary Material, Secs.~S3 and~S4.}
\vspace{-0.4cm}
\label{fig:Graph4}
\end{figure}

We now examine Eq.~\eqref{solutionMinNorm} more closely. The assumption that $A^{l}$ is sparse for all $l=1,\ldots,2f-1$ guarantees that $W_{f}$  
is also sparse. For example, for a  square lattice network with  $N=35^2$,
$f=5$, and identity $B$, matrix $W_{f}$ is (multi-) banded and its  fill-in
(percentage of nonzero entries)
 is $9.99\%$. 
However, 
$W_{f}^{-1}$ is a 
dense matrix (inversion ``destroys" the sparse matrix structure). Because $W_{f}^{-1}$ is dense, from the block row of Eq.~\eqref{solutionMinNorm} corresponding to $\overline{\mathbf{u}}_{i}^{f}$
we conclude that the control input sequence of node $i$ is a function of the initial and desired states of potentially all nodes in the network. In other words, Eq.~\eqref{solutionMinNorm} constitutes a centralized controller for the network.

At first, it might appear that our attempt to find an input sequence for node $i$  as a function of the states of only neighboring nodes is doomed to fail.  However,   even though $W^{-1}_{f}$ is fully populated, this matrix exhibits the important property of 
being \textit{localized}.   
This property is central to the solution we present.  
Localized matrices are characterized by having a relatively small number of entries that are (in absolute value) dominantly large compared to the others~\cite{Benzi2015,haber2014};
prominent  examples include (but are not limited to)
off-diagonally decaying matrices~\cite{demko1984,benzi2007,canuto2014decay}. 
To investigate how the degree of localization may depend on the condition number $\kappa$ of $W_{f}$, we first note that  $\kappa$ 
depends on the choice of 
$B$. To examine these dependencies, we consider
three block diagonal matrices $B$: 
$B_{1}=\text{diag}\big[B_{ii}^{(1)}\big]$, $B_{2}=\text{diag}\big[B_{ii}^{(2)}\big]$, and $B_{3}=\text{diag}\big[B_{ii}^{(3)}\big]$, where  
$B_{ii}^{(1)}=\text{diag}\,(1,1,1)$, $B_{ii}^{(2)}=\text{diag}\,(1,1,0)$, and $B_{ii}^{(3)}=\text{diag}\,(1,0,0)$  for $i=1,2,\ldots,N$.

Figure~\ref{fig:Graph4}(a-c) shows the 
absolute values of the entries of the inverse Gramians $W_f^{-1}=[z_{ij}]$ organized in decreasing order
 for an Erd\H{o}s-R\'enyi network (ERN), a random geometric network (RGN), and a square lattice network, respectively.
The different curves correspond to $B=B_{1}$, $B_{2}$, and $B_{3}$.
In each case, a small number of $|z_{ij}|$  values are significantly larger than the others, and the remaining values decrease 
faster than exponentially. This localization is more dominant for better-conditioned Gramians, especially for the  RGN 
and lattice network.
 (The structures in the $B_{3}$ curves result from unevenness internal to the $n\times n$ blocks of $W_f^{-1}$.)
These  
results are in accordance with theoretical
results on localization phenomena in the inverses of sparse symmetric matrices~\cite{Benzi2015} and demonstrate 
that many networks have localized $W_f^{-1}$.

To further analyze this property, we first 
consider the case of
lattice networks, which have banded matrices $A$ and $W_f$ for suitable node orderings. 
Because $W_f$ is positive definite,
from Ref.~\cite{demko1984} it follows that $W^{-1}_{f}$ is an off-diagonally decaying matrix. 
Specifically, the off-diagonal decay rate 
of $W^{-1}_{f}$ is 
given  by $|z_{ij}|\le c\lambda^{|i-j|}$, where   
\begin{equation}
c=\Big\|W^{-1}_{f} \Big\|_{2}\max \Big\{1, \frac{\left(\sqrt{\kappa} +1\right)^{2}}{2\kappa}  \Big\}, \,\, \lambda=\left[\frac{\sqrt{\kappa}-1}{\sqrt{\kappa}+1} \right]^{2/\beta},
\label{offDiagonalDecayofQExplained}
\end{equation}
and $\beta$ is the bandwidth of $W_{f}$~\cite{demko1984}. From Eq.~\eqref{offDiagonalDecayofQExplained} it follows that the off-diagonal decay rate of $W^{-1}_{f}$ is faster if the condition number
$\kappa$ is smaller, indicating that the inverses of well-conditioned Gramians have fast off-diagonal decay. 
The inset of Fig~\ref{fig:Graph4}(c) shows a confirmation of this prediction for a lattice network.
Relations between localization and condition number can also be established for other localized $W^{-1}_{f}$,
as we show explicitly in Supplementary Material, Sec.~S5, for general network structures.

The above inspires us to use the fundamental result that a localized matrix can be approximated by 
a sparse matrix~\cite{grote1997parallel,chow1998approximate,benzi2007,Benzi2015,haber2016}.
We thus propose a sparse approximation $X$ of the dense matrix $W^{-1}_{f}$. 
This  approximation can be found by solving the constrained least-squares problem
\begin{align}
\min_{X} \left\|I-W_{f}X \right\|_{F}^{2}, \,\, \text{subject to}\;  X\in\mathcal{X},
\label{leastSquaresInversion}
\end{align}
where $\mathcal{X}$ is the set of matrices with a given sparsity pattern (i.e., the pattern of nonzero entries)
and $\left\|\cdot \right\|_{F}$ denotes the Frobenius norm~\cite{grote1997parallel,chow1998approximate}.  In Eq.~\eqref{leastSquaresInversion} we have to choose the \textit{a priori} sparsity pattern of $X$, which determines the set $\mathcal{X}$. 
For a sparse matrix,
a good choice for the sparsity pattern of the inverse
is given by a sum of powers of the matrix itself~\cite{huckle1999apriori}.
This,  together with the fact that $W_{f}$ can be expressed in terms of the form  $A^{j}BB^{T}(A^{T})^{j}$, motivates 
us to choose the \textit{a priori}  sparsity pattern of $X$ as the sparsity pattern of
$\overline{X}_{q}=I+\sum_{l=0}^{q}A^{l}BB^{T}\left(A^{T} \right)^{l}$,
where the  integer $q\ge0$ is a user choice 
(further theoretical justification is offered in Supplementary Material, Sec.~S4).
Assuming that 
$\overline{X}_{q}$ is a
sparse matrix,  the computational and memory complexities of solving Eq.~\eqref{leastSquaresInversion}  scale linearly with the dimension $Nn$ (in contrast
with the quadratic complexities of sparse matrix inversion that would be required to determine $W_f^{-1}$~\cite{golub2012}). 
 Given that the  optimization problem in Eq.~\eqref{leastSquaresInversion}
is also highly parallelizable~\cite{grote1997parallel,chow1998approximate}, 
this enables us to compute $X$ for very large networks. 

Equipped with a sparse approximation $X$ of $W^{-1}_{f}$, we can now define a localized controller with associated SINs determined by the sparsity pattern of $X$.
Specifically, substituting $W^{-1}_{f}$ with $X$ in Eq.~\eqref{solutionMinNorm}, we obtain the approximate input sequence
$\widehat{\overline{\mathbf{u}}}^{f}=\mathcal{Q}\mathbf{x}^{d} -\mathcal{R}\mathbf{x}(0)$, 
where $\mathcal{Q}=\mathcal{K}^{T}_{f}X$ and $\mathcal{R}=\mathcal{K}^{T}_{f}X\mathcal{A}$. 
The input sequence of node $i$ can then be expressed as
\begin{align}
\widehat{\overline{\mathbf{u}}}_{i}^{f}=\sum_{j \in \mathcal{N}_{1}(i)} \mathcal{Q}_{ij} \mathbf{x}_{j}^{d}-\sum_{j \in \mathcal{N}_{2}(i)} \mathcal{R}_{ij} \mathbf{x}_{j}(0),
\label{localRelationShipInput}
\end{align} 
where $\mathcal{Q}= [\mathcal{Q}_{ij}]$, $\mathcal{R}=[\mathcal{R}_{ij}]$, and $\mathcal{N}_{1}(i)$ and $\mathcal{N}_{2}(i)$ are 
sets of indices defining the corresponding SINs of node $i$. 
We can show that the SINs of node $i$ are determined by the set of nodes 
that are up to a certain distance from $i$. 
For  $B=B_{1}$, in particular,  we obtain 
that for {\it any} undirected network, $\widehat{\overline{\mathbf{u}}}_{i}^{f}$ 
is a function of  $\mathbf{x}^{d}_{j}$ for nodes $j$ within distance
$2q+f-1$ from $i$ and is a function of  $\mathbf{x}_{j}(0)$ for nodes $j$ within distance
$2(q+f)-1$  from $i$ in the network (Supplementary Material, Sec.~S4.1). 
From the 
relation between powers of the adjacency matrix and walks in the network, 
these neighborhoods are guaranteed to be generally small relative to $N$ provided that 
$A^{\pm}_{q,f}=A^{2q+ (3\pm 1)f/2-1}$ are sparse; moreover, as explained next, Eq.~\eqref{localRelationShipInput} can define a localized controller even when this condition is violated.
We first note, however, that
a solution $X$ in Eq.~\eqref{leastSquaresInversion} can be found, and thus a suitable localized controller may still be defined,
even when $W_f$ is not invertible (i.e., 
when $f<\nu$, where  smaller $f$ leads to sparser $A^{\pm}_{q,f}$).
 
\begin{figure}[t] 
\centering
\includegraphics[scale=1,trim=0mm 0mm 0mm 0mm,clip=true]{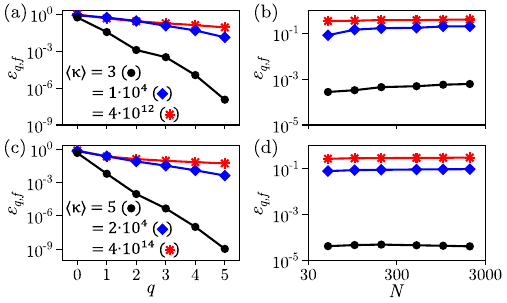} 
\caption{Control error for (a, b) ERNs and (c, d) RGNs
as a function of (a, c) $q$ and (b, d) $N$, where $B=B_1$ ($\bullet$), $B_2$ ($\scriptstyle\blacklozenge$), and $B_3$ ($\boldsymbol{*}$). 
The fixed parameters are $f=3$ in (a-d), $N=598$ in (a), $N=702$ in (c),  and  $q=2$ in (b, d)~\cite{param}.
Each data point is an average over $100$ realizations of matrix $A$ in (a, c) and 
of network topologies in (b, d).
For the sparsity patterns of the associated matrices $\mathcal{Q}$ and $\mathcal{R}$, see Supplementary Material, Sec.~S4.3.}  
\label{fig:Graph6}
\vspace{-0.4cm}
\end{figure}
 
An example of a SIN associated with $\mathcal{Q}$ in a lattice network 
is shown in 
Fig.~\ref{fig:Graph4}(d), 
where $\|\mathcal{Q}_{ij}\|_F$ 
decreases rapidly away from $i$. 
This rapid decay, which is common across the networks we consider, can be used to define
further reduced {\it effective} SINs by ignoring all nodes~$j$ in the SIN of node~$i$ for 
which $\|\mathcal{Q}_{ij}\|_F$ are smaller than a certain threshold $\delta$, as shown in Fig.~\ref{fig:Graph4}(d) for $\delta=10^{-6}$ and $10^{-3}$. The use of effective SINs, which are defined analogously for $\mathcal{R}$, 
can significantly reduce the
computational and communication costs.
For ERNs, for instance, 
matrices $A^{\pm}_{q,f}$ are generally not sparse,
which leads to larger SINs.
However, the effective SINs are still relatively small and lead only to a modest reduction in accuracy (Supplementary Material, Sec.~S4, 
where the sizes of the SINs and effective SINs are quantified by the fill-ins of the matrices $\mathcal{Q}$ and $\mathcal{R}$). 
This shows that our approach 
can be applied even when $A^{\pm}_{q,f}$ are not sparse.

The approximation involved in replacing 
$W_{f}^{-1}$ by $X$ could 
in principle 
reduce the performance of the localized controller~\eqref{localRelationShipInput}  compared to the centralized controller~\eqref{solutionMinNorm}. 
This approximation depends
on the size of the SINs in Eq.~\eqref{localRelationShipInput}, which are in turn modulated by our choice of $q$.
From our results above for localized $W_{f}^{-1}$, it follows that, for  well-conditioned $W_{f}$, 
the SINs of the localized controller can be very
small without significant loss of performance.
When $W_{f}$ is not well-conditioned, on the other hand,
 there should be a trade-off between the performance of the localized controller and the size of the SINs. 
To quantify this trade-off, we measure the performance 
of the localized controller 
using as a metric the relative distance from the desired state  
 $\varepsilon_{q, f} =\left\|\mathbf{x}^{d} -\widehat{\mathbf{x}}(f) \right\|_{2}/\left\|\mathbf{x}^{d}  -\mathbf{x}(0)  \right\|_{2}$, where $\widehat{\mathbf{x}}(f)$ is the right side of Eq.~\eqref{leastSquaresFormulation2} 
 for $\overline{\mathbf{u}}^{f}$ replaced by  $\widehat{\overline{\mathbf{u}}}^{f}$. 
Note that $\varepsilon_{q, f}$ measures the control error over all nodes in the network. 

\begin{figure}[t!] 
\includegraphics[scale=1,trim=0mm 2mm 0mm 0mm,clip=true]{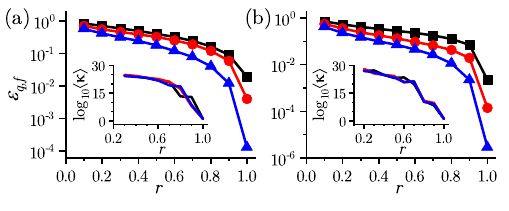} 
\caption{Control error versus the fraction of controlled nodes in (a) an ERN ($N=598$) and (b) an RGN ($N=702$), for $q=1$ ($\scriptscriptstyle\blacksquare$), $q=2$ ($\bullet$), and $q=3$ ($\scriptstyle\blacktriangle$).  
Each point is an average over $100$ realizations of randomly chosen nodes to control.
Insets: corresponding condition numbers of $W_f$ color-coded as in the main panels.  The other
settings are 
$f=5$ and  $B_{ii} = B_{ii}^{(1)}$ if node $i$ is controlled (and zero otherwise)~\cite{param}.}
\vspace{-0.4cm}
\label{fig4}
\end{figure}

\section{Illustrations of the Approach}

Figure~\ref{fig:Graph6} shows $\varepsilon_{q, f}$ for both ERNs and RGNs,
where $X$ is computed using the SPAI algorithm~\cite{grote1997parallel} to solve Eq.~\eqref{leastSquaresInversion}. 
Similar results are obtained for lattices (Supplementary Material, Sec.~S6).
As shown in Fig.~\ref{fig:Graph6}(a, c), the control performance
is excellent already for small $q$ (hence small SINs) and
further improves as  $q$  
is increased. 
For fixed SINs, the performance also improves as
$\kappa$ is decreased ($\langle\kappa\rangle$ indicates the average $\kappa$ for the realizations shown).  
Moreover,  as a function of $N$, the
error $\varepsilon_{q, f}$ does not vary significantly for large network sizes [Fig.~\ref{fig:Graph6}(b, d)]. The latter confirms the potential of the approach for efficient control of large networks. This potential becomes even more evident when communication constraints are accounted for, since small SINs imply that the communication complexity 
and delays for each node become independent of the network size, which further enhances the overall
performance of the localized approach  
(Supplementary Material, Sec.~S7).

The localized control approach is also effective when only a fraction $r$ of the nodes is controlled, as shown in 
Fig.~\ref{fig4} 
 for ERN and RGN systems.   
The 
error $\varepsilon_{q, f}$ is small  for relatively small $r$ and
 becomes significantly smaller as $r$ increases. 
 The
  increase in $r$ is accompanied by a significant decrease of  $\kappa$, and thus by 
 a higher localization in $W_{f}^{-1}$.
 Therefore,
 the relationship 
 between the condition number  and localization 
 also governs the trade-off between 
 control performance
 and the fraction of controlled nodes.

\begin{figure}[t!]   
\centering
\includegraphics[width = 3.3 in]{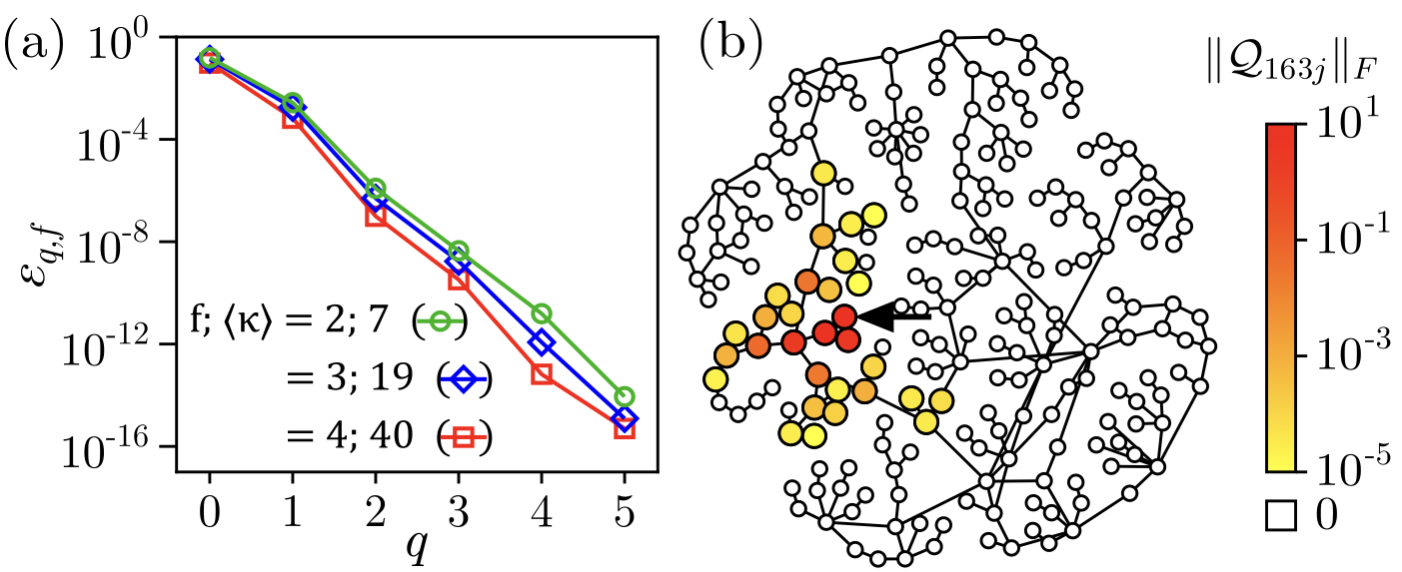} 
\caption{
Control of Iceland's power grid when driving the generators to the desired synchronous state.
(a) Control error  for several $f$ as a function of $q$. Each 
point is an average over $100$ realizations of $\mathbf{x}(0)$ generated by randomly perturbing each state variable by 
$\pm 0.01$ away from $\mathbf{x}^d$, and $B$ is the identity matrix.
(b) SIN of the node marked by the arrow for the matrix $\mathcal{Q}$ and $f=q=2$.  
Details of the dynamics are presented in Supplementary Material, Sec.~S9.
}
\label{fig:Graph62}
\vspace{-0.4cm}
\end{figure}

To demonstrate the applicability of the approach to real networks of significant interest, 
we consider the problem of controlling the synchronization of power generators in 
a power grid. We explicitly consider the Iceland's power grid~\cite{iceland-data}, which 
comprises 35 generators and 189 buses (nongenerator nodes). 
We focus on scenarios for which the synchronization of all generators at the required 
frequency---a condition for the system to operate---corresponds to an unstable state
and we seek to  control the system with this 
state serving as $\mathbf{x}^d$. 
This system is described by a nonlinear set of dynamical equations, with two state variables (phase 
and frequency) for each generator and one state variable (phase) for each bus.
Using our approach, we compute the control law for the dynamics linearized 
around  $\mathbf{x}^d$. 
Figure~\ref{fig:Graph62}(a) shows that we are indeed able to reach the desired target state efficiently;  
the scaling of $\varepsilon_{q, f}$ versus $q$ shows similar behavior to that of model 
networks (cf. Fig.~\ref{fig:Graph6}). Figure~\ref{fig:Graph62}(b)  shows for a typical node that  
 the resulting SINs remain generally small. 
Moreover, provided that the initial state is not too far from $\mathbf{x}^d$, the control actions defined by the linearized dynamics successfully drive the system to the desired state 
when applied directly to the full nonlinear dynamics,
which illustrates the usefulness of the approach
in the study of nonlinear systems 
(details are offered in  Supplementary Material, Secs.~S9 and~S10).

\section{Concluding Remarks}

In the localized control approach developed here, the control actions for a node are determined solely using state information in a neighborhood around the node. The resulting control is generally only approximate but the reduction in computational and communication complexity can vastly overcompensate the loss of accuracy by allowing more frequent control inputs. This provides a foundation for the use of our approach not only in networks for which the global state information is not readily available but also in networks for which the control actions would be too costly to compute using all the available state information.  
It follows that even very large networks can be controlled with efficiency and efficacy in a localized manner.

Our approach for linearized systems also serves as a foundation for developing a localized control methodology for nonlinear systems. For nonlinear systems, the equivalent formulation is feedforward model predictive control, where the control problem becomes a nonlinear optimization problem. While such optimization problems generally lack closed-form solutions, they can be solved numerically, often using variations of Newton's method. Arguments similar to those presented above can then be used to show once again that sparsity and localized inverses of control matrices imply localized control laws. Exploring this for nonlinear network systems is an important problem for future research.

We conclude by noting that there is now a significant demand for the local and computationally efficient implementation of control, state estimation, optimization, and machine-learning algorithms that can avoid or minimize long-distance data communication across networks. 
Thus, in addition to its immediate significance for open-loop control, our approach may serve as a fundamental building block for more complex algorithms. 

\section{Supplementary Material}
The online Supplementary Material presents further details, examples, analyses, and figures in support of the core results presented in this article. 
Additionally, the following GitHub repository includes images of the state information neighborhoods for all nodes in Iceland's power grid considered in Fig.~\ref{fig:Graph62} and the random geometric network in Fig.~S13: \url{https://github.com/fmolnar-notredame/network_control_paper/}.

\section{Acknowledgments}

A.E.M. thanks Prof. David K. Campbell---on the occasion of this Focus Issue to celebrate his career---for his many direct and indirect contributions as a pioneer and model scientist in nonlinear science and for his role in the creation of the journal {\it Chaos} and the Center for Nonlinear Studies.
This work was supported by Simons Foundation Award No.~342906, ARPA-E Award No.~DE-AR0000702, ARO Grant No.~W911NF-19-1-0383,  MURI ARO Grant No.~W911NF-14-1-0359, and MURI ARO Grant No.~W911NF-24-1-0228. 
The views and opinions of authors expressed herein do not necessarily state or reflect those of the United States Government or any agency thereof.

\section{AUTHOR DECLARATIONS}

\noindent
{\bf Author Contributions}
 
 \noindent
{\bf Aleksandar Haber:} Conceptualization (equal);  Formal analysis (lead); Investigation (lead); Software (equal); Visualization (equal);  Writing -- original  draft (lead);  Writing -- review \& editing (equal).
{\bf Ferenc Molnar:} Investigation (supporting); Formal analysis (supporting); Investigation (supporting); Software (equal); Visualization (equal); Writing -- original  draft (supporting); Writing -- review \& editing (equal).
{\bf Adilson E. Motter:} Conceptualization (equal);  Funding acquisition (lead);  Project administration (lead); Supervision (lead); Visualization (supporting); Writing -- original  draft (supporting); Writing -- review \& editing (equal).\\

\noindent
{\bf Conflict of Interest}

\noindent
The authors have no conflicts to disclose.\\

\noindent
{\bf Data Availability}

\noindent
The data that support the findings of this study are available
within the article, supplementary material, and associated GitHub
repository.


\begin{thebibliography}{99}
\bibitem{param}
In these simulations, we  use 
$\mathbf{x}(0)=(0,\dots,0)$ and $\mathbf{x}^{d}=(1,\dots,1)$.
\end{thebibliography}
\end{document}